\documentstyle[12pt]{article}
\begin{document}
\renewcommand{\thefootnote}{\fnsymbol{footnote}}

\title{{\bf Kalb-Ramond interaction for a closed p-brane}}
\author{Motowo Yamanobe\footnotemark[2], Sumio Ishikawa\footnotemark[3],
        Yasuhiro Iwama,\\
        Tadashi Miyazaki, Kazuyoshi Yamamoto
        and Ritsu Yoshida\footnotemark[1] \\
{\it Department of Physics, Science University of Tokyo} \\
{\it Kagurazaka, Shinjuku-ku, Tokyo 162, Japan}}
\footnotetext[1]{and Asia University Junior College}
\footnotetext[2]{E-mail: yamanobe@gradsun.ap.kagu.sut.ac.jp}
\footnotetext[3]{E-mail: liszt@gradsun.ap.kagu.sut.ac.jp}

\date{August 1994}

\maketitle
\vspace{20mm}
\begin{flushleft}
\small{
}
\end{flushleft}

\vspace{20mm}
\begin{abstract}
The Kalb-Ramond action for an interacting string
is generalized to the case of a high-dimensional object (p-brane).
The interaction is found to be mediated by a gauge boson of a completely
antisymmetric tensor of rank $p+1$.
\end{abstract}

\setlength{\baselineskip}{8mm}
\newpage
The superstring theory was proposed as one of the candidates for the Theory of
Everything (TOE).
It had been obtaining brilliant success until the end of 1980s,
when it was found to include some difficulties,
such as those about physical dimensions and about degeneracy of vacuum states.
Above all, the superstring theory tells nothing about the question
why the extended object must be one-dimensional.
Are there any possibility that a high-dimensional one be a candidate for
the TOE?
Along this line the attempt to extend an object of a one-dimensional string
to a p-brane (p:an arbitrary positive integer) has been done.
In order to recognize the p-brane theory with the aim to complete the TOE,
we must elucidate it in all its aspects;
for example, the classical solutions regulating the p-brane itself,
its interactions and so on.
In case of $p=2$ (membrane),
some investigations on its solutions expressing a free membrane
have been done\cite{poppe},
but very little is known about the influence of interactions for the theory.
The main purpose of the present paper
is, therefore, to realize an interacting p-brane through the
action-at-a-distance (AD) force.

Long time ago,
Feynman and Wheeler showed that
a charged particle interacting with an electromagnetic field
can be described by the action-at-a-distance\cite{fey}.
Kalb and Ramond extended their idea on a point particle to that on
a one-dimensional extended object (string).
They obtained a gauge boson as mediating the
interaction between strings\cite{kalb}.
In case of a closed string,
their gauge boson is an antisymmetric tensor
of rank two and its degree of freedom is one.
Namely, it becomes a massless scalar meson.

The main purpose of this paper is to generalize Kalb-Ramond's
formulation to the the case of a {\it closed p-brane theory}.
Let us consider an extended object of {\it p} dimensions (p-brane)
which traces out $D$-dimensional space-time.
Here $X_a^\mu(\xi_a^i)$ represents such a p-brane,
where $\xi_a^i\,\,(i=0 \sim p)$
are the parameters needed to represent the world sheet,
and the subscript $a$ labels each p-brane.
The volume element of its world sheet is
%1
\begin{eqnarray}
d\sigma_a^{\mu_0\mu_1...\mu_p}=d^{p+1}\!\xi_a
\sigma_a^{\mu_0\mu_1...\mu_p},
\end{eqnarray}
where
%2
\begin{eqnarray}
\sigma_a^{\mu_0\mu_1...\mu_p}=
\frac{\partial
       \left(
         X_a^{\mu_0},X_a^{\mu_1},...,X_a^{\mu_p}
       \right)}
     {\partial
       \left(
         \xi_a^0,\xi_a^1,...,\xi_a^p
       \right)}.
\end{eqnarray}
The action for a free p-brane is given by
%3
\begin{eqnarray}
S_a^{free}=
-\int_{\xi_a^0=\tau_i}^{\xi_a^0=\tau_f}\!\!
\int_{\xi_a^1=0}^{\xi_a^1=a_1}\!\!
\dots
\int_{\xi_a^p=0}^{\xi_a^p=a_p}
  \left(
    -d\sigma_a\cdot d\sigma_a
  \right)^\frac{1}{2}.
\end{eqnarray}
In case of a {\it closed} p-brane, with a boundary condition $(a_i>0)$
\[X_a^\mu(\xi_a^0,0,0,...,0)=X_a^\mu(\xi_a^0,a_1,a_2,...,a_p),\]
we obtain the equation of motion, by imposing that
the action (3) be stationary under the variation
%4
\begin{eqnarray}
X_a^\mu(\xi_a^i) &\rightarrow &
X_a^\mu(\xi_a^i)+\delta X_a^\mu(\xi_a^i), \nonumber \\
\delta X_a^\mu(\tau_i,\xi_a^1,...,\xi_a^p) &=&
\delta X_a^\mu(\tau_f,\xi_a^1,...,\xi_a^p)=0,
\end{eqnarray}
%5
\begin{eqnarray}
(p+1)D_a^{\mu_1\mu_2...\mu_p}
  \left[
    \frac{\sigma_{a\,\,\mu\mu_1\mu_2...\mu_p}}
         {(-\sigma_a \cdot \sigma_a)^{\frac{1}{2}}}
  \right]
=0,
\end{eqnarray}
where
%6
\begin{eqnarray}
D_a^{\mu_1\mu_2...\mu_p} &=& \sum_{i=0}^p
\frac{\partial \sigma_a^{\mu\mu_1...\mu_p}}
     {\partial X_{a\, ,\,i}^\mu}
\frac{\partial}
     {\partial \xi_a^i}, \nonumber \\
X_{a\, ,\, i}^{\mu_j} &=& \frac{\partial X_a^{\mu_j}}
                               {\partial \xi_a^i}.
\end{eqnarray}

Now we take up two closed p-branes $a$ and $b$ interacting with each other
through AD forces,
and start with the following action
%7
\begin{eqnarray}
S=S_a^{free}+S_b^{free}+\int\!\!d^{p+1}\!\xi_a
\int\!\!d^{p+1}\!\xi_b R_{ab}(X_a,X_b,X_{a\,,\,i},X_{b\,,\,j}),
\end{eqnarray}
with
%8
\begin{eqnarray}
R_{ab}=R_{ba},
\end{eqnarray}
for symmetry requirement.
Imposing the condition that the action (7) be
stationary under the variation (4),
the equation of motion is obtained as follows:
%9
\begin{eqnarray}
(p+1)D_a^{\mu_1\mu_2...\mu_p}
  \left[
    \frac{\sigma_{a\,\,\mu\mu_1\mu_2...\mu_p}}
         {(-\sigma_a \cdot \sigma_a)^{\frac{1}{2}}}
  \right]
=\int\!\!d^{p+1}\!\xi_b
  \left(
    \frac{\partial R_{ab}}
         {\partial X_a^\mu}
    -
    \frac{\partial}
         {\partial \xi_a^i}
    \frac{\partial R_{ab}}
         {\partial X_{a\,,\,i}^\mu}
  \right).
\end{eqnarray}
Multiplying Eq.(8) by $X_{a\,,\,i}^\mu\,\,(i=0 \sim p)$,
we immediately have
%10
\begin{eqnarray}
0=
  \left[
    \frac{\partial}{\partial \xi_a^i}
      \left(
        1-\sigma_a^{\mu_0\mu_1...\mu_p}
        \frac{\partial}{\partial \sigma_a^{\mu_0\mu_1...\mu_p}}
      \right)
  \right]
\int\!\!d^{p+1}\!\xi_bR_{ab},
\end{eqnarray}
which shows that
$R_{ab}$ must be a linear function of the $\sigma_a^{\mu_0\mu_1...\mu_p}$.
Equation (10) can also be obtained by demanding that the action be
invariant under the reparametrization
%11
\begin{eqnarray}
\xi_a^i \rightarrow \xi_a^i+\delta \xi_a^i.
\end{eqnarray}

Assuming that $R_{ab}$ depends on $X_{a\,,\,i}^\mu$ only through
the combinations of $\sigma_a$ and $\sigma_b$,
Eq. (9) is rewritten as
%12
\begin{eqnarray}
(p+1)D_a^{\mu_1\mu_2...\mu_p}\!\!
  \left[
    \frac{\sigma_{a\,\,\mu\mu_1\mu_2...\mu_p}}
         {(-\sigma_a \cdot \sigma_a)^{\frac{1}{2}}}
  \right]\!\!
=\!\!\int\!\!d^{p+1}\!\xi_b\!\!
  \left(
    \frac{\partial R_{ab}}{\partial X_a^\mu}
    -(p+1)D_a^{\mu_1\mu_2...\mu_p}
    \frac{\partial R_{ab}}{\partial \sigma_a^{\mu\mu_1...\mu_p}}
  \right).
\end{eqnarray}
We explicitly take the following form as the interaction $R_{ab}$,
%13
\begin{eqnarray}
R_{ab}=g_ag_b\sigma_a^{\mu_0\mu_1...\mu_p}\sigma_{a\,\mu_0\mu_1...\mu_p}
       G(S_{ab}^2),
\end{eqnarray}
with
\[S^2_{ab}=(X_a-X_b)\cdot(X_a-X_b),\]
where $g_a$ and $g_b$ are coupling constants
and $G$ is some Green's function in $D$-dimensional space-time,
representing time-symmetric interactions,
i.e., composed of two parts;
advanced and retarded parts.
In this case the equation of motion is reduced to the following,
%14
\begin{eqnarray}
(p+1)D_a^{\mu_1\mu_2...\mu_p}\!\!
  \left[
    \frac{\sigma_{a\,\,\mu\mu_1\mu_2...\mu_p}}
         {(-\sigma_a \cdot \sigma_a)^{\frac{1}{2}}}
  \right]
=g_aF_b^{\mu\mu_0...\mu_p}(X_a)\sigma_{a\,\mu_0\mu_1...\mu_p}.
\end{eqnarray}
Note that Eq.(14) is much similar to the equation describing a
charged particle under the Lorentz force.
The analog of the electromagnetic field tensor is completely antisymmetric
 of rank $p+2$, and it can be written as
%15
\begin{eqnarray}
F_b^{\mu \mu_0 \mu_1 \cdots \mu_p}(X)
&\equiv& \partial^\mu \phi_b^{\mu_0 \mu_1 \cdots \mu_p}(X)\nonumber \\
& & +(-1)^{|p| \cdot 1} \partial^{\mu_0}
\phi_b^{\mu_1 \mu_2 \cdots \mu}(X) \nonumber \\
& & +(-1)^{|p| \cdot 2} \partial^{\mu_1}
\phi_b^{\mu_2 \mu_3 \cdots \mu_0 }(X) \nonumber \\
& & \vdots \nonumber \\
& & +(-1)^{|p| \cdot (p+1)} \partial^{\mu_p}
\phi_b^{\mu \mu_1 \cdots \mu_{p-1}}(X),
\end{eqnarray}
with
\[
|p|=0 \,\,\,\,{\rm for}\,\,p:{\rm odd},
\]
and
\[|p|=1 \,\,\,\, {\rm for}\,\,p:{\rm even},
\]
where $\phi_b^{\mu_0\mu_1...\mu_p}(X)$ is the tensor potential construct
due to string $b$ of rank $p+1$,
%16
\begin{eqnarray}
\phi_b^{\mu_0\mu_1...\mu_p}(X) &=&
 g_b\int\!\!d^{p+1}\!\xi_b\sigma_b^{\mu_0\mu_1...\mu_p}G((X-X_b)^2).
\end{eqnarray}
It is found, by an explicit calculation, that
%17
\begin{eqnarray}
\partial_{\mu_0}\phi_b^{\mu_0\mu_1...\mu_p}(X)=0.
\end{eqnarray}
$G(X)$ is the general Green's function in $D$ dimensional space-time,
and obeys the equation
%18
\begin{eqnarray}
\left(
  \partial^\mu\partial_\mu+m^2
\right)
G(X^2)=-C\delta^{(D)}(X),
\end{eqnarray}
where $C$ is a dimensionless constant and
$m$ is to be identified with a mass of the field.
By Eq. (18),
we have
%19
\begin{eqnarray}
\left(
  \partial^\mu\partial_\mu+m^2
\right)
\phi_b^{\mu_0\mu_1...\mu_p}(X)=
-Cj_b^{\mu_0\mu_1...\mu_p}(X),
\end{eqnarray}
where $j_b^{\mu_0\mu_1...\mu_p}(X)$ is the matter current,
%20
\begin{eqnarray}
j_b^{\mu_0\mu_1...\mu_p}(X)=
g_b\int\!\!d\sigma_b^{\mu_0\mu_1...\mu_p}\delta^{(D)}(X-X_b).
\end{eqnarray}
These equations (18)-(20) express a gauge field
mediating the interacting p-branes with one another.

On the analogy of the case of the string\cite{kalb},
we will transfer to the field theory.
The gauge field only couples to the p-brane through the field tensor (15),
and the coupling is invariant under the gauge transformation
%21
\begin{eqnarray}
\phi^{\mu_0 \mu_1 \cdots \mu_p} \rightarrow
& & \phi^{\mu_0 \mu_1 \cdots \mu_p} \nonumber  \\
& & +\partial^{\mu_0} \Lambda^{\mu_1 \mu_2 \cdots \mu_p }(X)
\nonumber \\
& & +(-1)^{|p| \cdot 1} \partial^{\mu_1}
\Lambda^{\mu_2 \mu_3 \cdots \mu_0}(X) \nonumber \\
& & +(-1)^{|p| \cdot 2} \partial^{\mu_2}
\Lambda^{\mu_3 \mu_4 \cdots \mu_1 }(X) \nonumber \\
& & \vdots \nonumber \\
& & +(-1)^{|p| \cdot p} \partial^{\mu_p}
\Lambda^{\mu_0 \mu_1 \cdots \mu_{p-1}}(X),
\end{eqnarray}
where $\Lambda^{\mu_1 \mu_2 \cdots \mu_p}$ is completely antisymmetric
with respect to its indices.
The Lagrangian density invariant by the transformation (21),
from which the equation of motion (14) is led,
is given by
%22
\begin{eqnarray}
{\cal L}(X)=
\frac{(-1)^{|p|}}{2 \cdot(p+2)!} F^{\mu \mu_0 \mu_1 \cdots \mu_p}(X)
F_{\mu \mu_0 \mu_1 \cdots \mu_p}(X).
\end{eqnarray}
The field $\phi^{\mu_0\mu_1...\mu_p}$ is massless.
The total action for the closed p-brane is, therefore,
%23
\begin{eqnarray}
S=-\int\!\! \sqrt{-d\sigma\!\cdot\!d\sigma}
+g\int\!\!d\sigma\!\cdot\!\phi +\int\!\!d^D\!X{\cal L},
\end{eqnarray}
with ${\cal L}$ given by Eq.(22).
We immediately find that
the degrees of freedom for the field $\phi^{\mu_0\mu_1...\mu_p}$ are
%24
\begin{eqnarray}
\left\{
\begin{array}{ll}
{}_{D-2}C_{p+1} &\,\, {\rm for}\,\,D \geq p+3, \\
\,\,\,\,\,\,\,\,\,\,0 & \,\,{\rm for}\,\,D=p+1,\,p+2.
\end{array}
\right.
\end{eqnarray}

\vspace{8mm}
In conclusion, we have obtained
the action for an interacting closed p-brane.
The interaction term $R_{ab}$ must be, and, in fact, {\it is}
a linear function of the $\sigma_a^{\mu_0\mu_1...\mu_p}$.
One might have other interactions with
the above linearity\cite{ramond}.
The equation of motion for the p-brane in our system is similar to
that for a charged particle under the Lorentz force.
By Eqs.(18)-(20) we see that the field
$\phi^{\mu_0\mu_1...\mu_p}$ is a gauge boson.
It corresponds to the photon in the electromagnetic theory and its degrees
of freedom are calculated to be ${}_{D-2}C_{p+1}$.

In this paper, we have confined ourselves to {\it closed} p-branes.
The investigation on the interacting {\it open} p-branes
will be reported shortly.

\vspace{8mm}
One of the authors (M.Y.) thanks K.Hoshi for a valuable discussion.
Two of the authors (K.Y. and M.Y.) would like to thank
Iwanami F\^{u}jukai for financial support.

\end{document}